\documentclass[11pt,fleqn]{article}
\usepackage{amsfonts}
\usepackage{graphicx}
\usepackage{float}
\restylefloat{figure}
\usepackage{cite}
\title{\bf  
Fermi's golden rule and exponential decay as a RG fixed point}
\date{\today}
\author{
Edwin Langmann\thanks{\tt E-mail: langmann@kth.se}\,
and G\"oran Lindblad\thanks{\tt E-mail: gli@kth.se}
 \vspace{.5cm} \\
{\normalsize  Department of Theoretical Physics}\\
{\normalsize  Royal Institute of Technology}\\
{\normalsize  SE-106 91 Stockholm, Sweden} 
 }
\columnwidth\textwidth
\renewcommand{\Re}{\hbox{\rm Re}\,}
\renewcommand{\Im}{\hbox{\rm Im}\,}
\def\C{{\mathbb C}}
\def\R{{\mathbb R}} 
\def\Z{{\mathbb Z}} 
\def\half{\frac{1}{2}\,}

\def\uni{{\hbox{{\rm 1}\kern-.24em\hbox{\rm l}}}}
\begin{document}
\maketitle
\begin{abstract}
We discuss the decay of unstable states into a quasicontinuum
using models of the effective Hamiltonian type.
The goal is to show that exponential decay and the golden 
rule are exact in a suitable scaling limit, and 
that there is an associated renormalization group (RG) with 
these properties as a fixed point.
The method is inspired by a limit theorem for infinitely divisible
distributions in probability theory, where there is a RG with a 
Cauchy distribution, i.e. a  Lorentz line shape, as a fixed point.
Our method of solving for the spectrum is well known;  it
does  not involve a perturbation expansion in the interaction,
and needs no assumption of a weak interaction.
We  use random matrices  for the interaction, 
and show that the ensemble fluctuations vanish in the scaling limit.
Thus the limit is the same for every model in the 
ensemble with probability one. 
\end{abstract}

\section{Introduction}

The standard textbook derivation of Fermi's golden rule  
starts from a perturbation expansion of the unitary evolution, 
keeping the lowest nontrivial order, and then sums over a
dense set of final states, to get a decay or reaction rate 
for an unstable state
\begin{equation}
\Gamma = 2 \pi \rho \,v_{}^{2}  
\label{gr}
\end{equation}
Here $v$ represents the (average) transition matrix element
and $\rho$ is the density of final states.
(We  will use the  convention $\hbar = 1$ throughout.)
The formula is essentially contained in Dirac \cite{dirac27},
see also \cite{weisskopf30}. 
If the rate $\Gamma$  is a constant  there results an 
exponential decay of the occupation number 
$ p(t) = \exp (- \Gamma t ) p(0)$.
 The corresponding quantum amplitudes are the Fourier transforms 
of a  Lorentz line shape function, see equation (\ref{lorentz}). 

It has been known for a long time that there are models where 
the golden rule and the exponential decay can be obtained without
a perturbation expansion.
It is our goal to show that such results hold for a  
large ensemble of models, and are exact in a suitable limit which 
leaves $\Gamma$ invariant.
This paper treats only the mathematical aspects of the models,
mainly using known methods.
It does not attempt to justify their physical relevance, but 
there are numerous applications of such models to lineshape
problems, quantum chemistry, and many other fields; 
we will give a few references later.

The approach we use here was inspired by an important theorem of 
probability theory. 
The  Lorentz line shape function (\ref{lorentz}) is  a  Cauchy 
probability density (PD), an infinitely divisible stable distribution.
It is the distribution of a properly scaled limit of an 
infinite sum of independent identically distributed random variables,
see Feller's book \cite{feller}. 
For any initial distribution the limit depends only on two 
real parameters,  $(\Gamma , a)$ in (\ref{lorentz}). 
This is similar to the role of the normal distributions
in the central limit theorem, but the limit is adapted to PDs 
with long tails. 

A physicist would call the Cauchy PD a fixed point for
a renormalization group (RG) of transformations on the space of 
PDs, see Sections \ref{cauchy} and \ref{anexample}. 
Our thesis is that this is also a good way to approach 
our problem. 
We will define an ensemble of Hamiltonians where the spectral
density defining the decay is a closed expression  (\ref{imr1})
in the Hamiltonian. 
There is no need for a perturbation expansion!
In a simple scaling limit this spectral density will converge 
to one of the Cauchy type, a limit  
that can also be interpreted as the fixed point of a RG.

Many attempts at a rigorous derivation of rate equations (and 
quantum Markov master equations) use the van Hove limit, where 
there is a scaling of the interaction matrix elements $(v)$ 
and the time $(t)$  as follows
 \begin{equation}
v \rightarrow \lambda v, 
\quad t \rightarrow  \lambda_{}^{-2}t 
\label{wscale}
\end{equation} 
where  $\lambda \rightarrow 0$ 
\cite{vanhove54, vanhove57,davies76,derezinski06}. 
In this way the dimensionless quantity $\Gamma t $ is invariant, 
but after the scaling the decay is on the rescaled (slow) time.
Before taking such a limit it is necessary to let the 
density of states (DOS) $\rho $ be infinite, otherwise there would be 
no relaxation in the limit $\lambda \rightarrow 0$. 
Here we will instead use the scaling 
 \begin{equation}
v \rightarrow \lambda v, 
\quad \rho \rightarrow  \lambda_{}^{-2} \rho
\label{cscale}
\end{equation} 
where $\lambda \rightarrow 0$,  $\Gamma $ is invariant,
 and $t$ does not scale. 
The name continuum, or statistical, limit is 
often used, but note that here all eigenstates are 
normalized and the spectrum is always discrete  (a quasicontinuum). 
The dimensionless number 
\begin{equation}
N_{\Gamma}^{} := \rho \,\Gamma 
\label{ngamma}
\end{equation}
scales as 
$N_{\Gamma}^{} \rightarrow \lambda_{}^{-2} N_{\Gamma}^{}$, 
and has a natural interpretation as the number  of states under
the resonance width.
It is also the number of matrix elements effectively involved in the 
transition,  a natural large parameter for the limit (\ref{cscale}).
We could replace the small parameter $\lambda $
by $N_{\Gamma}^{-1/2}$.

We will show that (\ref{cscale}) is the relevant scaling both for the 
limit theorem for the Cauchy PD, and for the quantum models 
introduced in Section \ref{matrixmodels}, and that the scaling can be 
interpreted as the iteration of a RG.
In the scaling limit the models will have exponential relaxation,
and the rate will be given by the golden rule (\ref{gr}) or a 
matrix generalization of this formula.

An outline of the contents is as follows. 

We first discuss the probabilistic properties of the Cauchy
PD in Sections \ref{cauchy} and \ref{anexample}.
In Section \ref{matrixmodels} we introduce the models of the 
the effective Hamiltonian type and the spectral density for 
the relaxation is found, without any kind of expansion or 
approximation.
The models are defined as an ensemble of random matrices; 
in the simplest case the ensemble is defined by just two variable 
(non-random) parameters $\rho$ and $v$. 

The simplest case, a single level decaying into a quasicontinuum,
is covered in Section \ref{simplemodel}.
For mathematical simplicity we perform the calculations for the case 
where the unperturbed spectrum lies in 
a finite energy interval $\Delta $ with energy range $\Delta E$.
We can easily calculate the ensemble averages and show that they 
have the desired properties in  the scaling limit.
When we let  $\Delta \rightarrow \R $ 
the $\lambda = 0 $ limit of the spectral density is a Lorentzian
with relaxation rate $\Gamma$ defined by (\ref{gr}).

In Section \ref{fluctuations} the variance of the fluctuations 
around the ensemble averages are estimated and shown to 
converge to zero in the scaling limit. 
Using the Chebyshev inequality we then conclude that 
each model in the ensemble will, with probability one, have an 
exponential relaxation with rate $\Gamma$ in this limit. 
We also get a measure of how good this is as an approximation 
for finite parameter values; 
this enhances the physical relevance of the models.
In particular, if  $N_{\Gamma}^{}\gg 1$ and $\Gamma \ll \Delta E$, 
the exponential relaxation is still a good approximation. 
We also note that the statistical ensembles can be dispensed 
with and replaced by certain uniformity assumptions on the 
spectra and interaction matrix elements.

In Section \ref{scaling} the relation between the scaling 
(\ref{cscale}) and  the RG for the matrix models is analyzed.
In Section  \ref{modifiedmodel} the unitary invariance of the 
random matrix ensemble is used to show that the results 
will hold for more general forms of the interaction.
This invariance also allows us to show how the properties of 
the Cauchy distribution under convolution are related to those of the 
spectral density.
In Section \ref{lineshape} we sketch how the models can
describe several decaying states, and the relation to Fano's 
theory of lineshape is pointed out.
Section \ref{timescales} spells out the limits to exponential decay
posed by a finite spectral range or  a discrete spectrum.
Finally there are some conclusions and additional remarks
in Section \ref{discussion}.

\section{Properties of the Cauchy distribution}
\label{cauchy}

The Lorentz (Breit-Wigner) line shape functions form a 
family of Cauchy PDs on $\R$, with parameters  $a \in \R$ and
$\Gamma > 0$
\begin{eqnarray}
f_{C}^{}(x - a, \Gamma) & = & 
\frac{\Gamma}{2 \pi} \:
\frac{1}{(x - a)_{}^{2} + \Gamma_{}^{2}/4}
\label{lorentz} \\ 
\int_{\R}^{} dx \,f_{C}^{}(x, \Gamma) & = & 1 
\nonumber 
\end{eqnarray}
Note that the mean  and variance  are not defined!
The distribution function for $a = 0$ is
\begin{equation}
F_{C}^{}(y, \Gamma) = \int_{-\infty}^{y} dx \, f_{C}^{}(x, \Gamma)
= \half + \frac{1}{\pi} \arctan 
\left[ \frac{2 \, y}{\Gamma}\right]
\label{cdistr}
\end{equation} 
When $\Gamma \rightarrow 0$ the limit is a unit step
function, $F_{C}^{} = 0$  for $y < 0$, and $F_{C}^{} = 1$ 
for $y > 0$. Of course 
$F_{C}^{}(- \infty, \Gamma) = 0$ and 
$F_{C}^{}(\infty, \Gamma) = 1$  for all $\Gamma \geq 0$.
The functions (\ref{lorentz}) form a convolution semigroup  
\begin{equation}
\int_{\R}^{} dx f_{C}^{}(x - a_{1}^{}, \Gamma_{1}^{})  
f_{C}^{}(y - a_{2}^{}  - x, \Gamma_{2}^{})
= f_{C}^{}(y - a_{1}^{} - a_{2}^{},\Gamma_{1}^{}  + \Gamma_{2}^{} )
\label{cconv}
\end{equation} 
see  \cite{feller} section 2.4.
Thus, if two independent RVs $X_{1}^{}$  and $X_{2}^{}$
have Cauchy PDs with parameters $\Gamma_{1}^{}$ and
$\Gamma_{2}^{}$, then $X_{1}^{}+ X_{2}^{}$ has a Cauchy PD
with parameter $\Gamma_{1}^{} + \Gamma_{2}^{}$.
For $N$ independent RVs 
$\{X _{1}^{},X _{2}^{}, \ldots , X _{N}^{}\}$, 
all with the same Cauchy PD with parameter $\Gamma$,  the sum 
is Cauchy with parameter $N \Gamma$
and the average is Cauchy with parameter $\Gamma$
\begin{equation}
\frac{1}{N} S_{N}^{} = 
\frac{1}{N} \sum_{k=1}^{N} \, X _{k}^{} 
\sim X _{1}^{}
\label{scale1}
\end{equation} 
where  $\sim$ indicates equality in distribution.
The characteristic function (CF) is 
\begin{equation}
\chi( t) := \int_{\R}^{} dx \, f_{C}^{}(x - a, \Gamma)
\exp (- i x t) =
\exp \left [- i a t  -   \half \Gamma\, \vert t \vert  \right ]
\label{charf}
\end{equation} 
The convolution (\ref{cconv}) corresponds to a multiplication 
of the CFs.

Let the RVs in (\ref{scale1}) have a common distribution $F$.
Even when $F$ is not Cauchy the scaled sum (\ref{scale1})
can converge to a limit 
\begin{equation}
X_{\infty}^{}:= \lim_{N \rightarrow \infty}^{}
\frac{1}{N} \,S_{N}^{} 
\label{clim}
\end{equation} 
with a Cauchy distribution.
Sufficient conditions on $F$ for convergence can be found in 
 \cite{feller} Section 17.5, especially in the Concluding remark, 
involving an assumption on the ``tails'' of $F$. 
The simplest case is that the following two limits exist and 
are equal
\begin{equation}
\lim_{x \rightarrow \infty}^{} x \left [1 - F(x)\right] =
\lim_{x \rightarrow \infty}^{} x F(-x) = 
\frac{\Gamma}{2 \pi}  
\label{tails}
\end{equation} 
where we can already identify the Cauchy parameter $\Gamma$.
We let $\chi_{F}^{}$ denote the CF of $F$ 
and define a sequence of centering constants (with a
dimensional  parameter $\gamma$)
\begin{equation}
\beta_{k}^{} := \gamma \int_{\R}^{}  \sin (x/k\gamma) F(dx), \quad 
 k \in \Z_{+}^{}
\label{center}
\end{equation} 
It  follows from (\ref{tails}) that the following limit exists 
\cite{feller}
\begin{equation}
\lim_{n \rightarrow \infty}^{}
 \left [ \chi_{F}^{} (t/n)e_{}^{i \beta_{n}^{}t}
\right ]_{}^{n} = e_{}^{- \vert t \vert \Gamma /2}
\label{limcfun}
\end{equation} 
The statement for the  distribution $F$ corresponding to 
(\ref{limcfun}) is
\begin{equation}
F_{\infty}^{} (x ):= 
\lim_{n \rightarrow \infty}^{} (\ast_{}^{n}\,F )
 [n (x  + \beta_{n}^{})] = F_{C}^{}( x, \Gamma )
\label{limitf}
\end{equation} 

We note the similarity  to the central limit theorem, but there 
the factor $1/N$ in (\ref{clim}) is replaced by  $1/\sqrt{N}$. 
If $ F $ has zero mean and  a finite second moment then 
there is a convergence to a normal distribution, see \cite{feller}, 
Section 8.4. 
A RG version of the central limit theorem is described in  
\cite{jonalasinio01}.
On the other hand for an $ F $ with  
a finite second moment the limit (\ref{limitf}) is a step
function corresponding to a  $\delta$-function density.

\section{An example}
\label{anexample}

We now want to give an intuitive hint why the mathematical 
result in the previous section is connected to the quantum 
lineshape problem. 
As an example of a distribution which fulfills (\ref{tails})
we pick  one which is obtained from a standard lowest order 
perturbation expansion. 
Consider an infinite set of unperturbed quantum states 
$\{\vert k \rangle, \, k \in \Z\}$, 
with energy eigenvalues 
$  E_{k}^{} = k \omega $. 
There is a perturbation    $V$  with nonzero matrix elements 
$\langle 0 \vert V \vert k\rangle = v, \, \forall k \neq 0$.
The perturbed state $\vert \psi_{0}^{}\rangle$ which converges to 
 $\vert 0 \rangle $ as $v \rightarrow 0$
has the occupation numbers, to the order $v_{}^{2}$
 \begin{equation}
p_{k}^{} := \vert \langle k \vert \psi_{0}^{}\rangle \vert_{}^{2} 
=  \frac{v_{}^{2}}{k_{}^{2} \omega_{}^{2}}
\; (k\neq 0),
\quad 
p_{0}^{} :=  1 - 2 \sum_{k=1}^{\infty} p_{k}^{}  
\label{pk}
\end{equation} 
where $v_{}^{2}$ must be small enough to make $p_{0}^{} > 0$,
i.e.   
$v_{}^{2}  \leq 3 \pi_{}^{- 2}\,\omega_{}^{2}$.
 The distribution is a step function with
steps at $x = k \omega$
\[
F(x) = \sum_{k \omega < x}^{}p_{k}^{}
\]
and centered, hence (\ref{center}) is zero.
We find that the first limit in (\ref{tails}) reads 
\[
\frac{v_{}^{2}}{\omega_{}^{2}}
 \lim_{x \rightarrow \infty}^{}  x 
\sum_{k \omega \geq x}^{} \frac{1}{k_{}^{2}}
= \frac{v_{}^{2}}{\omega_{}^{2}}
 \lim_{x \rightarrow \infty}^{} x 
\int_{y \omega \geq x}^{} \frac{dy}{y_{}^{2}}
= \frac{v_{}^{2}}{\omega} = \rho v_{}^{2}
= \frac{ \Gamma}{ 2 \pi}
\]
and the second limit is identical; consequently we know that 
(\ref{limitf}) holds with $\beta_{n}^{} = 0$.
 
Note that  $F \ast F$ has steps separated by $\omega$, while 
$(F \ast F) [ 2 x ]$ 
has steps separated by $\omega/2$, and  $(\ast_{}^{n} F)(n x)$ 
in (\ref{limitf}) has steps separated by $\omega/n$. 
On the other hand we know from (\ref{cconv}) if $F$ is Cauchy 
then this convolution and rescaling recovers $F$.
Consequently, this is an example of the scaling 
(\ref{cscale}), when we identify $\Gamma$ with (\ref{gr}).

As a preparation for  later developments we consider 
a generalization to random distributions.
Let $V$ have  random  matrix elements 
$\langle 0 \vert V \vert k \rangle = \xi_{k}^{}$
 where the RVs $\xi_{j}^{}$ are assumed to be complex-valued, normal, 
and independent, defined by the first two moments 
\begin{equation}
\left \langle  \xi_{k}^{}\right \rangle = 0, 
\quad  \left \langle \xi_{j}^{\ast} \xi_{k}^{} \right\rangle
= v_{}^{2}\delta_{j k}^{} 
\label{xirand1}
\end{equation}
The distribution function is now itself a random function.
Consider the distribution of the tail function
\[
T(x) := 1 - F(x) = \frac{1}{ \omega_{}^{2}}
\sum_{k \omega \geq x}^{} 
\frac{\vert \xi_{k}^{} \vert_{}^{2} }{k_{}^{2}}
\]
Clearly the ensemble average gives back what we had before 
\[
 \lim_{x \rightarrow \infty}^{} x \left  \langle T(x) \right \rangle 
= \frac{ \Gamma}{ 2 \pi}
\]
We can also calculate the ensemble variance. 
Introduce the real random vector
 \begin{equation}
\eta_{k}^{} := \vert \xi_{k}^{} \vert_{}^{2} - v_{}^{2}
\label{eta1}
\end{equation} 
and use the standard properties of the normal RVs $\xi$
to calculate
 \begin{equation}
\langle \eta_{k}^{} \rangle =  0, \quad 
\langle \eta_{j}^{} \eta_{k}^{} \rangle = \delta_{j k}^{}\: v_{}^{4}
\label{eta2}
\end{equation} 
Define
\[
\Delta T(x) := T(x) - \left \langle T(x) \right\rangle 
= \frac{1}{ \omega_{}^{2}}
\sum_{k \omega \geq x}^{} 
\frac{  \eta_{k}^{} }{k_{}^{2}}
\]
Then $\langle \Delta T(x)\rangle = 0$ and for large $x$
\[
\langle \Delta T(x)_{}^{2} \rangle 
=  \frac{1}{ \omega_{}^{4}}
\sum_{k \omega \geq x}^{} 
\frac{ v_{}^{4} }{k_{}^{4}} \approx
 \frac{ v_{}^{4}}{ \omega_{}^{4}}
\int_{y \omega \geq x}^{} \frac{dy}{y_{}^{4}}
= \frac{  v_{}^{4}}{ 3\,\omega_{}^{} x_{}^{3}}
\]
and it follows that 
\[
\lim_{x \rightarrow \infty}^{}
x_{}^{2}\,\langle \Delta T(x)_{}^{2} \rangle 
= \lim_{x \rightarrow \infty}^{}
\frac{  v_{}^{4}}{ 3\,\omega_{}^{} x } = 0
\]
Using the Chebyshev inequality we find as $x \rightarrow \infty$
(see \cite{feller}, Chapter 5)
\[
{\rm Probability} \left \{  \vert x \Delta T (x) 
\vert \ge \delta \right \}
\leq \delta_{}^{-2}
x_{}^{2}\,\left \langle \Delta T(x)_{}^{2} \right\rangle 
\approx \frac{v_{}^{4}}{3\, \delta_{}^{2} \omega}
\frac{1}{x} \rightarrow 0
\]
We conclude that with probability 1 an element 
 $F$ in the ensemble will satisfy (\ref{tails}), 
consequently 
  (\ref{limcfun}) and (\ref{limitf}) will hold.
However, the centering constants (\ref{center})
can not be left out when the $\xi_{k}^{}$ have the 
distribution defined by (\ref{xirand1}).

\section{Matrix models and resolvents}
\label{matrixmodels}

In this section we introduce the matrix models of the 
effective Hamiltonian type.
For most of the discussion we assume the Hilbert space 
to be of finite dimension. 
In  the limit $\lambda \rightarrow 0$ in (\ref{cscale})   the 
dimension will be infinite, but again we emphasize that the 
eigenstates are all normalized.  
Also the limit of an unbounded discrete spectrum with finite 
DOS is interesting.
We assume that these limits can be performed on the final 
results of the calculations  without
dealing too deeply with problems of mathematical rigor.

We note that the formulas derived in this section are known, 
the main ideas going back at least to Feshbach 
\cite{feshbach58,feshbach62, lowdin62}. 
The mathematical background can be traced from 
Remark 2.1 of \cite{jensen06}.
There are also textbook treatments with physical applications, 
 see \cite{cohentannoudji92}, Complements  
$C_{I}^{}$ and $C_{III}^{}$, and \cite{messiah}, Chapter 21.
However, we will use the results in a way which seems not to be 
standard.

Consider a self-adjoint matrix $H$ (Hamiltonian), with discrete spectrum 
$\{\omega_{\nu}^{}\}$, assumed non-degenerate for simplicity,
and spectral projectors $\{P_{\nu}^{}\}$
\[
H := \sum_{\nu}^{} \omega_{\nu}^{}P_{\nu}^{}
\]
Define a  causal resolvent (Green's function) 
with a regularization parameter  $\varepsilon > 0$
\[
R (z - i \varepsilon) := 
(z - i\varepsilon - H)_{}^{-1}
= \sum_{\nu}^{} (z -  i\varepsilon  - \omega_{\nu}^{})_{}^{-1}
P_{\nu}^{} 
\]
We can pick a number of the projectors by integrating over a  
counterclockwise contour encircling  the poles with real parts  
in a finite interval $I \subset \R$:
$\{ z = \omega_{\nu}^{} + i \varepsilon,\: \omega_{\nu}^{} \in I\}$ 
\[
\sum_{\omega_{\nu}^{} \in I}^{} P_{\nu}^{} 
= \frac{1}{2 \pi i}\oint dz \: R(z - i \varepsilon)
\]
For small $\varepsilon$ we can approximate the contour integral 
by one along the real axis
\[
\frac{1}{2 \pi i} \int_{I}^{} dx \,[\, R (x  - i\varepsilon ) - 
 R (x + i\varepsilon) ]
= \frac{1}{ \pi }\int_{I}^{} dx  \:\Im R (x - i\varepsilon)
\]
This means that we can consider the function 
\begin{equation}
\frac{1}{ \pi }\,\Im R (x - i\varepsilon)
= \frac{1}{ \pi } \sum_{\nu}^{} 
\frac{\varepsilon}{ (x -\omega_{\nu}^{})_{}^{2}
+\varepsilon_{}^{2} } \: P_{\nu}^{} 
\label{spdensity}
\end{equation} 
as a regularized spectral density,  normalized as follows
\[
\frac{1}{ \pi } \int_{\R}^{} dx\, \Im R (x - i\varepsilon)
= \sum_{\nu}^{} P_{\nu}^{}  = \uni
\] 
and the trace of  (\ref{spdensity})
is a regularized DOS.
The regularization can be written as a convolution (averaging)
by a Cauchy PD
\[
\Im R (x - i\varepsilon) = \lim_{\delta \rightarrow 0}^{}
\int_{\R}^{} dy \, f_{C}^{}(x-y,\varepsilon) \:
\Im R (y - i \delta)
\]
The regularization can be justified by noting that that the duration
of an observation of the system is bounded by a time scale 
$1/\varepsilon$, assumed much longer than the decay time.

Now write $H$ as a  block matrix, 
with projectors $\Pi_{A}^{}, \Pi_{B}^{}= \uni - \Pi_{A}^{}$ 
on the complementary subspaces
\begin{equation}
H = \left [ 
\begin{array}{c c}
H_{A}^{} & V \\
V_{}^{\dagger} & H_{B}^{}
\end{array}
\right]
\label{twopart}
\end{equation} 
We know from the properties of the Schur complement 
\cite{cottle74} that  
 \begin{equation}
R_{A}^{}(x - i\varepsilon) :=
\Pi_{A}^{}  R (x - i\varepsilon ) \Pi_{A}^{} = 
[ x - i \varepsilon - \tilde{H}_{A}^{}(x - i\varepsilon) ]_{}^{-1}
\label{resa}
\end{equation} 
where,  for any $z \in \C$, we define a non-Hermitian effective
Hamiltonian
\begin{equation}
\tilde{H}_{A}^{}(z) : = 
H_{A}^{}  +  V (z - H_{B}^{} )_{}^{-1} V_{}^{\dagger}
= \tilde{H}_{A}^{}(z_{}^{\ast})_{}^{\dag}
\label{htilde}
\end{equation} 
We find  for the imaginary part of (\ref{resa}), which, with 
a factor $1/\pi$, is the  spectral density  (\ref{spdensity}) projected 
on the subspace $A$  
 \begin{eqnarray}
\phi_{A}^{}(x,\varepsilon) &:= &\frac{1}{\pi}\, 
\Im R_{A}^{}(x - i\varepsilon) \nonumber
 \\
& = & \frac{1}{\pi}\,  R_{A}^{}(x - i\varepsilon) \,
\left [\, \varepsilon + \Im \tilde{H}_{A}^{}(x - i\varepsilon)
\right ] \,
 R_{A}^{}(x - i\varepsilon)_{}^{\dagger} 
\label{imr1}
\end{eqnarray} 
For all $\varepsilon > 0, \; x \in \R $ this is a positive definite matrix.
The FT is the matrix-valued CF 
\begin{equation}
\chi_{A}^{}(t, \varepsilon) := 
  \int_{\R}^{} dx\, \exp (- i x t) \,
\phi_{A}^{}(x,\varepsilon) 
\label{char2}
\end{equation}
We also define the distribution function 
\begin{equation}
\Phi_{A}^{}(x,\varepsilon):= 
\int_{-\infty}^{x} dy\, \phi_{A}^{}(y,\varepsilon)
\label{distr2}
\end{equation}
Note that the normalization implies that for every $\varepsilon$, 
\[
\Phi_{A}^{}(-\infty,\varepsilon) = 0, \quad
\Phi_{A}^{}(\infty,\varepsilon) = \uni_{A}^{}
\]
When $\varepsilon \rightarrow 0$  (\ref{distr2}) converges to a 
step function $\Phi_{A}^{}(x, 0)$; the steps are at
$x = \omega_{\nu}^{}$, the spectrum of $H$.

We can see  in (\ref{imr1}) the beginning of an exponential 
relaxation.
In fact, if we could assume that 
$\tilde{H}_{A}^{}(x - i\varepsilon)$
is independent of $x$, then (\ref{char2}) is a matrix-valued
decaying amplitude
 \begin{equation}
\chi_{A}^{}(t, \varepsilon)
= \exp \,\left ( - i \, t \,\Re \tilde{H}_{A}^{} 
 - \vert t \vert \varepsilon 
- \vert t \vert \Im \tilde{H}_{A}^{} \right)
\label{ft}
\end{equation} 
Our goal is to justify this simple form in the limit (\ref{cscale}),
where we can finally set $\varepsilon = 0$.
In most of the calculations below we will let the
subspace  $A$ have rank 1, and then, if we leave out the 
argument $ x - i\varepsilon$, 
the projected spectral density simplifies to 
\begin{equation}
\phi_{A}^{}(x,\varepsilon)  
= \frac{1}{\pi}\,\frac{\varepsilon  
+ \Im \tilde{H}_{A}^{}(x - i\varepsilon)}
{(x - \Re \tilde{H}_{A}^{}(x - i\varepsilon))_{}^{2}
+ (\varepsilon + \Im \tilde{H}_{A}^{}(x - i\varepsilon))_{}^{2}}
\label{imr}
\end{equation} 
Again, if $\tilde{H}_{A}^{}(x - i\varepsilon)$ is constant we 
have the resonance form (\ref{lorentz}).

It is important for the success of this approach that the 
$\varepsilon$-averaged form of the resolvent (\ref{imr1})
is expressed in terms of $\Im \tilde{H}_{A}^{}(x - i\varepsilon) $
which is also an $\varepsilon$-averaged expression, 
as we will see below.

\section{The  simplest model}
\label{simplemodel}

The model (\ref{twopart}) is particularly simple when the
subspace indexed $A$ has rank 1, and the spectral density 
has the form (\ref{imr}).
This case has been solved repeatedly
since a long time, and used for numerous applications, 
see for example 
\cite{fano61, bixon68, freed76, cohentannoudji92, dietz07}.
Here  the interaction $V$ will represent an ensemble of random 
matrices.
This is a device often used in applications \cite{carmeli80, mehta},
here it allows us to make statements valid for almost all 
elements in the ensemble.

Now $H_{A}^{}$ is represented by single state 
$ \vert \psi_{s}^{}\rangle $, of energy $E_{s}^{} $, while 
 $H_{B}^{}$ is spanned by a finite set of states 
 $\{ \vert j \rangle; j = -N_{B}^{}, - N_{B}^{}+1, \ldots , N_{B}^{}\} $. 
First we choose an equidistant spectrum  
$E_{j}^{} = E_{0}^{} + j \, \omega_{B}^{}$ where
$E_{0}^{} \approx E_{s}^{}$.
The calculations will show that we can cope with more general spectra
with an average DOS $ \rho_{B}^{} = 1/\omega_{B}^{}$.
We also assume that the energies are restricted to a finite interval 
$\Delta$ of width $\Delta E = E_{+}^{} - E_{-}^{} $ 
\[
E_{j}^{} \in \Delta := [ E_{-}^{} , E_{+}^{} ]  
\]
Only a lower bound is essential for finite quantum systems,
but here we prefer to simplify the mathematics by having a 
finite number of states in $H_{B}^{}$, 
$N_{B}^{} = 1 + \rho_{B}^{}\, \Delta E$; 
later we can let this number approach $\infty$.
 
The interaction $V$ couples $ \vert \psi_{s}^{}\rangle $ 
with all the $\vert j \rangle$, allowing this state to decay.
We choose the matrix elements to be independent 
complex normal random variables 
 \begin{equation}
\langle \psi_{s}^{} \vert V  \vert j \rangle 
= \langle j \vert V  \vert  \psi_{s}^{} \rangle_{}^{\ast} =
\xi_{j}^{} 
\label{vrand1} 
\end{equation} 
defined by (\ref{xirand1}),
while all other matrix elements are zero.
Thus, in (\ref{twopart}),  $H_{A}^{} = E_{s}^{}$, 
$V$ is a row vector with components
$\xi_{j}^{}$, and $ H_{B}^{}$ is a diagonal matrix with elements
$E_{j}^{}$.
The spectral density  (\ref{imr}) is now
 \begin{equation}
\frac{1}{\pi} \langle \psi_{s}^{} \vert 
\, \Im R(x - i\varepsilon )\, \vert \psi_{s}^{} \rangle
= \frac{\varepsilon }{\pi} \sum_{\nu}^{} 
\frac{\vert \langle \psi_{s}^{}\vert \omega_{\nu}^{}\rangle \vert_{}^{2 }}
{( x - \omega_{\nu}^{})_{}^{2 } + \varepsilon_{}^{2 }}
\label{imr0}
\end{equation} 
and in the limit  $\varepsilon \rightarrow 0$
the FT (\ref{char2})  is the amplitude
 for staying in the  state  $\vert\psi_{s}^{} \rangle $:
\[
\langle \psi_{s}^{} \vert \exp (- i H t )
\vert \psi_{s}^{} \rangle
= \sum_{\nu}^{} \vert \langle \psi_{s}^{}
\vert \omega_{\nu}^{}\rangle \vert_{}^{2}
\exp ( - i t \omega_{\nu}^{})
\]
For given $E_{s}^{}$, $H_{B}^{}$ and $V$
the equation system to solve for the eigenvalues $\omega_{\nu}^{}$ 
and eigenvectors $\vert \omega_{\nu}^{}\rangle$ is  \cite{bixon68}
\begin{eqnarray*}
(E_{s}^{}- \omega ) \langle \psi_{s}^{}\vert \omega 
\rangle + \sum_{j}^{} \xi_{j}^{} \langle j \vert\omega  \rangle
 &= & 0\\ 
(E_{k}^{}- \omega ) \langle k\vert \omega \rangle + 
\xi_{k}^{\ast} \langle \psi_{s}^{} \vert\omega \rangle
 &= &   0
\end{eqnarray*}
We know that each eigenvalue $\omega_{\nu}^{}$ is located between
two unperturbed $(V = 0)$ eigenvalues, 
but we do not need the exact values.
Instead we can estimate the terms in the RHS of (\ref{imr}).
Decompose (\ref{htilde})
into real and imaginary parts, for insertion in (\ref{imr}).
 \begin{equation}
\tilde{H}_{A}^{}(x - i\varepsilon) = E_{s}^{}  + 
\sum_{j}^{} \frac{\vert \xi_{j}^{}\vert_{}^{2} ( x - E_{j}^{})}
{(x - E_{j}^{})_{}^{2} + \varepsilon_{}^{2}}
+ i \varepsilon \sum_{j}^{} \frac{\vert \xi_{j}^{}\vert_{}^{2}  }
{(x - E_{j}^{})_{}^{2} + \varepsilon_{}^{2}}
\label{htilde2}
\end{equation} 
We first deal with the imaginary part
 \begin{equation}
\Im \tilde{H}_{A}^{}(x - i\varepsilon) = 
\varepsilon \sum_{j}^{} \frac{\vert \xi_{j}^{}\vert_{}^{2}  }
{(x - E_{j}^{})_{}^{2} + \varepsilon_{}^{2}}
\label{imh}
\end{equation} 
A simpler case is solved e.g.\ in \cite{bixon68}, where the 
interaction is non-random, 
 $\xi_{j}^{} = v$, and the spectrum of $H_{B}^{}$
is unbounded, i.e. $E_{\pm }^{} = \pm \infty$.
When $\omega_{B}^{} \ll \varepsilon$ the infinite sum 
is approximated by an integral which is
independent of the argument in $\tilde{H}_{A}^{}$
 \begin{equation}
\Im \tilde{H}_{A}^{} \approx
\frac{\varepsilon v_{}^{2} }{\omega_{B}^{}}
\int_{\R}^{} dy \frac{1}{ y_{}^{2} +  \varepsilon_{}^{2} }
=  \pi \rho_{B}^{} v_{}^{2}  = 
\frac{\Gamma}{2}  
\label{imh2}
\end{equation} 
where we have put $\rho =  \rho_{B}^{}$ in (\ref{gr}).
For our model, first apply the ensemble average over the 
RVs $\xi_{j}^{}$, then, again, approximate by an integral
\begin{eqnarray}
\left \langle \Im \tilde{H}_{A}^{}  \right \rangle 
&= & \varepsilon v_{}^{2}\sum_{j}^{} \frac{1 }
{(x - E_{j}^{})_{}^{2} + \varepsilon_{}^{2}}
\approx 
\frac{\Gamma}{2} \, J(x, \varepsilon)
\label{avimh}
\\ 
J(x, \varepsilon)& : = & \frac{  \varepsilon}{\pi}
 \int_{E_{-}^{} -x}^{E_{+}^{} -x } dy \:
\frac{1}{ y_{}^{2} +  \varepsilon_{}^{2} }
= \arctan \left [ \frac{E_{+}^{} - x }{\varepsilon}\right ]
- \arctan \left [ \frac{E_{-}^{} - x }{\varepsilon}\right ]
\nonumber
\end{eqnarray}
It is easiest to see the behavior of the  function  $J$ by 
a computer calculation. 
For $\varepsilon \ll \Delta E$ it is close to the
CF  $\chi(\Delta, x)$  for the interval $\Delta$,
with deviations $\varepsilon$-near the end points.
When $\varepsilon \rightarrow 0$ then  
$J(x, \varepsilon) \rightarrow  \chi(\Delta, x)$ 
and 
\begin{equation}
\left \langle \Im \tilde{H}_{A}^{}(x - i\varepsilon) \right \rangle
\approx J(x, \varepsilon)\,\frac{\Gamma}{2}
\rightarrow   \chi(\Delta, x) \:\frac{\Gamma}{2}
\label{avimh2}
\end{equation}
The same method applies to the real part
\begin{eqnarray}
\left \langle \Re \tilde{H}_{A}^{} \right \rangle
&=& E_{s}^{} + 
v_{}^{2}
\sum_{j}^{} \frac{ ( x - E_{j}^{})}
{(x - E_{j}^{})_{}^{2} + \varepsilon_{}^{2}}
\approx E_{s}^{} + \Gamma K(x, \varepsilon)
\label{avreh}
\\ 
K(x,\varepsilon) & := &  \frac{1}{2 \pi}
\int_{E_{-}^{} - x }^{E_{+}^{} - x }dy\, 
\frac{y}{y_{}^{2} + \varepsilon_{}^{2}}
= \frac{1}{4 \pi}   \ln \left [ 
\frac{ (E_{+}^{} - x)_{}^{2} + \varepsilon_{}^{2}}
{ (E_{-}^{} - x)_{}^{2} + \varepsilon_{}^{2}} \right ]
\nonumber
\end{eqnarray}
The second term in (\ref{avreh}) represents an $x$-dependent 
level shift; it has a well-defined limit when 
$\varepsilon \rightarrow 0$.
It has the effect of changing the DOS slightly, on the order 
of $\Gamma / \Delta E$.
Near  the center $\bar{x}= (E_{+}^{} + E_{-}^{})/2$  of $\Delta$  
and for large  $\Delta E$
 \begin{equation}
K(x, 0) = \frac{1}{2 \pi} \ln \left [ 
\frac{ \Delta E - 2 (x - \bar{x})}
{ \Delta E + 2 (x - \bar{x})} \right ] 
\approx  - \frac{2}{\pi} \frac{x - \bar{x}}{\Delta E} 
\label{kx}
\end{equation} 
There is no unique way of taking the limit
$\Delta E \rightarrow \infty$, but as long as  $x$  and 
$\bar{x}$ stay bounded, it  holds for all fixed 
$\varepsilon \geq 0$ that
 \begin{equation}
\lim_{\Delta E \rightarrow \infty}^{}K(x, \varepsilon)  = 0
\label{limk}
\end{equation}

Summing up the results so far: if we use the ensemble averages 
derived above for the terms in (\ref{imr}), and replacing 
the sums with integrals, we find the following approximation
which we expect to be good for $\omega_{B}^{} \ll \varepsilon$
(cf. \cite{cohentannoudji92}, Complement $C_{III}^{}$)
\begin{equation}
\phi_{A}^{}(x, \varepsilon) \approx 
\frac{1}{2 \pi} \frac{2 \varepsilon +\Gamma J(x,\varepsilon)}
{[ x - E_{s}^{} - \Gamma K(x, \varepsilon) ]_{}^{2}
 + [ 2 \varepsilon + \Gamma J(x,\varepsilon)]_{}^{2}/4}
\label{imr2}
\end{equation}

\section{Estimating the ensemble fluctuations}
\label{fluctuations}

We want  to justify the calculations in the previous section
by showing that the ensemble fluctuations and other error
terms vanish in the limit (\ref{cscale}), thus that
$\Im\tilde{H}_{A}^{} $ and $\Re \tilde{H}_{A}^{} $ are given by
the last terms of  (\ref{avimh}) and  (\ref{avreh}),
and that (\ref{imr2}) becomes exact.
We will concentrate on  $\Im\tilde{H}_{A}^{}$, 
since the real part is very similar. 

The deviation from the ensemble mean  is given by
\[
\Delta \, \Im \tilde{H}_{A}^{} := 
\Im \tilde{H}_{A}^{}  - \left \langle \Im \tilde{H}_{A}^{}
\right \rangle 
= \varepsilon \sum_{j}^{} \frac{ \eta_{j}^{}  }
{(x - E_{j}^{})_{}^{2} + \varepsilon_{}^{2}}
\]
where $\eta$  is the RV (\ref{eta1}), with mean 
and variance given by (\ref{eta2}).
The variance of (\ref{imh}) is 
\[
\sigma_{}^{2}\left( \Im \tilde{H}_{A}^{}\right) :=
\left \langle \left (\Delta\,  
\Im \tilde{H}_{A}^{}  \right)_{}^{2} \right \rangle
= \varepsilon_{}^{2} v_{}^{4 }
\sum_{j}^{} \left [ (x - E_{j}^{})_{}^{2} + 
\varepsilon_{}^{2}\right]_{}^{-2}
\]
For an order of magnitude estimate we again replace the 
sum by an integral, and extend the integral to $\R$. 
Calculations similar to those performed above give 
 \begin{equation}
\sigma_{}^{2}\left( \Im \tilde{H}_{A}^{}\right) 
\approx \frac{\pi \,v_{}^{4}}{2 \varepsilon \,\omega_{B}^{}}
= \frac{\omega_{B}^{}}{2 \varepsilon} \frac{\Gamma_{}^{2}}{4 \pi}
\label{varh}
\end{equation} 
For the relative size of the fluctuations we can take 
the square root of this expression over (\ref{imh2}), 
estimating the importance of the correction due to 
these fluctuations  by the dimensionless parameter 
\begin{equation}
\kappa := \sqrt{\frac{\omega_{B}^{}}{2 \pi \varepsilon}}
=   \frac{1}{ \sqrt{ 2 \pi N_{\varepsilon}^{} }}
\label{corr1}
\end{equation} 
where $N_{\varepsilon}^{} := \rho_{B}^{}  \varepsilon$.
From  (\ref{cscale}) follows that it scales as
 $\kappa \rightarrow \lambda \kappa$ and goes to zero.
Similar calculations show that $\kappa$ also measures 
the fluctuations in 
$\Re \tilde{H}_{A}^{}$.
We again use the Chebyshev inequality for an upper bound on the 
probability of having a deviation from the 
ensemble average
\cite{feller} 
\[
\mathrm{Probability}\{ \vert \Delta \,\Im \tilde{H}_{A}^{} 
\vert \geq  \delta \}
\leq \frac{1}{2 \pi N_{\varepsilon}^{} }
\left [\frac{\Gamma}{2 \delta} \right ]_{}^{2}
\]
Clearly, this quantity scales as $\lambda_{}^{2}$, and   the 
scaling limit will give a value for $\Im \tilde{H}_{A}^{}$
and $\Re \tilde{H}_{A}^{}$ equal to the ensemble average.

We also have to deal with the error involved in 
replacing the sum over the spectrum by an integral, 
and in the assumptions on the level spacing. 
If the statistics for the level spacings form a Poisson process 
with parameter $\rho_{B}^{} $, then the expected number of 
levels in an interval $\delta E$ and the variance are equal
\[
\left \langle N_{\delta E}^{} \right \rangle = 
\rho_{B}^{} \delta E = \sigma_{}^{2}(N) :=
\left\langle (N_{\delta E}^{} -
\rho_{B}^{} \delta E)_{}^{2}\right \rangle
\] 
Using the random version of the sequence $\{E_{j}^{}\}$
in a sum like (\ref{htilde2}),  we get the 
relative size of the resulting fluctuations by setting 
$\delta E = \varepsilon$  
 \begin{equation}
\frac{\sigma_{}^{2}(N)}{\langle N \rangle_{}^{2}}
= \frac{1}{\rho_{B}^{}\, \varepsilon}
= \frac{\omega_{B}^{}}{\varepsilon }
\label{discerror}
\end{equation} 
The square root is of the same order as $\kappa$  (\ref{corr1}).
If instead we use a uniformly spaced spectrum, the integral 
approximation has a  smaller error,  of order (\ref{discerror})
squared.
We know that random matrix spectra are typically more uniform 
than the Poisson case; this is the feature 
of  ``spectral rigidity'' due to level repulsion \cite{mehta}. 
The correction term will then be in between the uniform and 
Poisson values. 
It appears reasonable to consider the Poisson value
as a worst case for ensemble of spectra with a given density 
for the ensemble average. 

The argument so far justifies (\ref{imr2}) as an exact
result  in the limit (\ref{cscale}).
What happens when the limits $\varepsilon \rightarrow 0$
and $\Delta \rightarrow \R$ are taken after (\ref{cscale})?
The order of taking the limits is not essential here.
The limit $\varepsilon \rightarrow 0$ is straightforward
in (\ref{imr2}), and we get a normalized
density $\phi_{A}^{}(x, 0)$. 
This is a modified Cauchy density in a finite 
spectral interval and with a level shift function from 
(\ref{avreh}) and (\ref{kx}).
Finally, in the limit $\Delta \rightarrow \R$, using (\ref{limk}),
we recover the exact Cauchy form
\begin{equation}
\lim_{\Delta \rightarrow \R}^{}
\phi_{A}^{}(x, 0) = f_{C}^{}( x - E_{s}^{}, \Gamma)
\label{lim1}
\end{equation}

\begin{figure}[H]
\center
\includegraphics{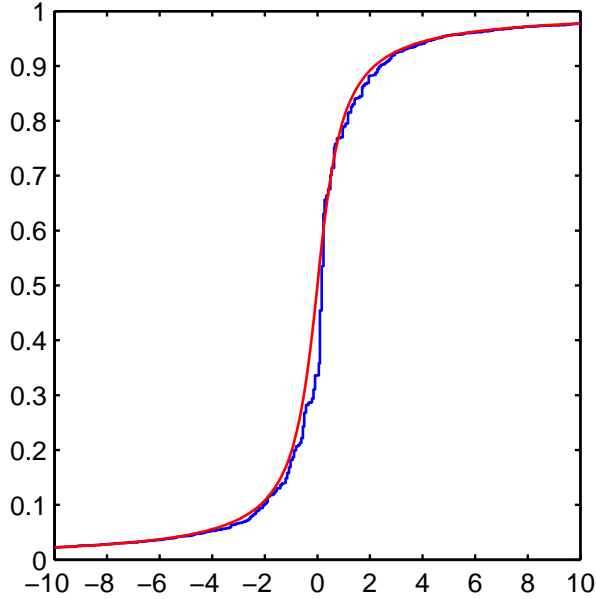}
\caption{\label{fig1}
Illustration of the closeness of the computed distribution 
function (the step graph) to the Cauchy form given by
(\ref{cdistr}) (the smooth graph) for a single random choice of the 
interaction $V$ in the model of Section \ref{simplemodel}.
The parameters are $N = 300$, $\Delta E = 20$,  $\Gamma = 1.41$.
Then $N_{\Gamma}^{} = 21$, not a very large number in this context. 
For larger values we see a convergence to the Cauchy form.
}
\end{figure}

In view of the fact that many applications involve models 
with large but finite DOS $\rho_{B}^{}$, it is also interesting 
to see what we can say in this case, when 
we let $\varepsilon \rightarrow 0$ while 
$N_{\Gamma}^{} \gg 1$.
It is then convenient to use the integrated distribution function 
(\ref{distr2}), which is always finite and monotonically increasing,
while $\phi_{A}^{}$ is  a sum of $\delta$-functions.
Thus $\Phi_{A}^{}(x, 0) $ is a step function, while the 
$\varepsilon$-averaged version 
$\Phi_{A}^{}(x, \varepsilon) $
is close, as measured by the small number (\ref{corr1}), 
to a Cauchy distribution (\ref{cdistr}) for 
$1 \ll N_{\varepsilon}^{} \ll N_{\Gamma}^{} $ 
and $\Gamma \ll \Delta E$.
Given the properties of these two functions, the local averaging on 
an energy scale  $\varepsilon$ cannot have a drastic effect, and 
it must hold that
\begin{equation}
\Phi_{A}^{}(x, 0) \approx F_{C}^{}( x - E_{s}^{}, \Gamma)
\label{lim2}
\end{equation}
is a good approximation.
This statement is  supported by numerical calculations, see 
Figure \ref{fig1}.
We can conclude that the distribution calculated from our 
model is close to Cauchy if there is a clear separation 
of energy scales
\begin{equation}
\omega_{B}^{} \ll  \Gamma \ll \Delta E
\label{ineq}
\end{equation}
which also implies that $N_{\Gamma}^{} \gg 1$.

To reach these conclusions, the properties of the spectrum 
of $H_{B}^{}$  and of the matrix elements of $V$ were crucial.
So far we have assumed simple regular or random distributions 
to do our estimates. 
Similar calculations are possible
for more general sequences $\{E_{j}^{}\}$
and  $\{\xi_{k}^{}\}$  without an assumed statistical distribution.
Instead we can postulate a uniformity for the spectrum and 
the matrix elements, a property which could in principle be 
verified in a concrete model.
For the level spacing assume that 
the following estimate  holds uniformly in $x$, 
for all $\varepsilon \gg \omega_{B}^{} $
and with $\kappa $ defined by (\ref{corr1}), 
 \begin{equation}
\frac{\omega_{B}^{} \: \varepsilon}{\pi} 
\sum_{j}^{} \frac{1}{ (x - E_{j}^{})_{}^{2} 
+ \varepsilon_{}^{2}} = 1 + O(\kappa)
\label{uni1}
\end{equation} 
When $\varepsilon \rightarrow \infty$ 
(or $\varepsilon \rightarrow \Delta E$ ) the RHS must be 
essentially unity, and this fixes $\omega_{B}^{}$.
In the same way the  uniformity of the matrix elements means that
 \begin{equation}
\frac{ \omega_{B}^{} \: \varepsilon}{\pi v_{}^{2}} 
\sum_{j}^{} \frac{\vert \xi_{j}^{}\vert_{}^{2}}{ (x - E_{j}^{})_{}^{2} 
+ \varepsilon_{}^{2}} = 1 + O(\kappa)
\label{uni2}
\end{equation} 
where  $v_{}^{2}$ is defined through the limit
 $\varepsilon \rightarrow \infty$.
The factor before the sum is $2 \varepsilon / \Gamma$.

\section{Scaling and renormalization}
\label{scaling}

It is easy to  see how the scaling 
(\ref{cscale}) works in the models of the type (\ref{twopart}).
If we disregard mathematical rigor the simplest form is
obtained when the energy interval $\Delta = \R$.
Then $H$  in  (\ref{twopart}) transforms in a simple
way: for $0 < \lambda < 1$ 
 \begin{equation}
 \left [ 
\begin{array}{c c}
H_{A}^{} & V \\
V_{}^{\dagger} & H_{B}^{}
\end{array}
\right] 
\longrightarrow 
 \left [ 
\begin{array}{c c}
H_{A}^{} &  \lambda V \\
\lambda V_{}^{\dagger} & \lambda_{}^{2} H_{B}^{}
\end{array}
\right] 
\label{rg1}
\end{equation} 
Clearly these transformations form a semigroup, and we 
think of them as a RG.
We can let $H_{B}^{}$  and $V$ be random or not; it is 
enough that they fulfill the  uniformity properties
(\ref{uni1}) and (\ref{uni2}), which are preserved 
under the transformations.
Of course, there is no proper limit for (\ref{rg1}) as 
$\lambda \rightarrow 0$.
For any  initial choice of $H$ of the type indicated above
 there a limit for 
 $\Phi_{A}^{}$ of the Cauchy form, and this is then 
the ``fixed point'' of the RG (\ref{rg1}).

When  $\Delta = \R$ we  can also discuss the relation of the 
RG transformation  (\ref{rg1})  with 
that defining the Cauchy PD limit in Section 
\ref{anexample}. 
For large $x$ the tail function of (\ref{distr2}) is 
\[
T(x,\varepsilon  ):= 1 - \Phi_{A}^{}(x, \varepsilon) 
\approx 
\frac{1}{\pi} \int_{x}^{\infty}
dy \: y_{}^{-2} \left[ \varepsilon +
 \Im \tilde{H}_{A}^{}(y - i \varepsilon)\right]
\]
For simplicity, again assume that the spectrum of 
$H_{B}^{}$ is $\{ k \omega_{B}^{}, \, k \in \Z \}$.
Then a straightforward calculation gives 
\[
\lim_{\varepsilon \rightarrow 0}^{} T(x,\varepsilon  ) = 
\frac{1}{\omega_{B}^{2}}
\sum_{k \omega_{B}^{} \geq x}^{}
\frac{\vert \xi_{k}^{}\vert_{}^{2}}{k_{}^{2}}
\]
Comparing with the calculations in Section \ref{anexample}
we again find  the tail condition, with probability 1  
\[
\lim_{x \rightarrow \infty}^{} x\, T(x,0  )
= \frac{\Gamma }{2  \pi}
\]
Hence, setting $\Phi_{A}^{}(x, 0) = F (x)$ we find 
that (\ref{limcfun}) and (\ref{limitf}) hold.
The two limiting procedures, that applied to the quantum 
models and that applied to PDs, give the same limits 
for an infinite uniform spectrum.
The same result holds for a spectrum that is a stationary
Poisson process.
 
We would  like to define a RG transformation also when 
$\Delta$ is a finite interval.
What does it look like in the matrix form (\ref{twopart})?  
We find  a scaling (\ref{cscale}) with $\lambda = 1/\sqrt{2}$,
and which involves a doubling of the dimension of the matrices.
To do this explicitly, consider a matrix composed of three parts,
where the assumptions on  $H_{A}^{}$ and the other components 
are as in Section \ref{simplemodel} 
\begin{equation}
H_{}^{} = \left [ 
\begin{array}{c c c }
H_{A}^{} & V_{B}^{} & V_{C}^{}  \\
V_{B}^{\dagger} & H_{B}^{} & 0 \\
V_{C}^{\dagger}  & 0 & H_{C}^{}
\end{array}
\right]
\label{threepart}
\end{equation} 
We note the matrix identity
\[
[V_{B}^{} , V_{C}^{}] \:
\left [ 
\begin{array}{c c  }
H_{B}^{} &  0   \\
0 & H_{C}^{}  
\end{array}
\right]_{}^{-1} \:
[V_{B}^{} , V_{C}^{}]_{}^{\dagger}
= V_{B}^{} H_{B}^{-1} V_{B}^{\dagger}
+ V_{C}^{} H_{C}^{-1} V_{C}^{\dagger}
\]
and the corresponding decomposition of the effective Hamiltonian  
\[
\tilde{H}_{A}^{}(z)  = H_{A}^{} + 
V_{B}^{}(z - H_{B}^{})_{}^{-1}V_{B}^{\dagger} + 
V_{C}^{}(z - H_{C}^{})_{}^{-1}V_{C}^{\dagger}
\]
Thus the contributions of $B$ and $C$ to $\Im \tilde{H}_{A}^{}$
and to the decay of $A$ just add up.
Clearly $\rho_{B+C}^{}  = \rho_{B}^{} +\rho_{C}^{}$.
In order deal as simply as possible with the energy level
statistics, we can assume Poisson statistics for $H_{B}^{}$
and $H_{C}^{}$ with parameters $\rho_{B}^{}$ and $\rho_{C}^{}$.
For $B+C$ there will then be Poisson statistics with 
parameter $\rho_{B+C}^{}$.

We can then let the $\lambda = 1/\sqrt{2}$  transformation 
be represented by   
\begin{equation}
H_{B}^{} \rightarrow H_{B + B_{}^{\prime}}^{}, \quad 
V \rightarrow  \frac{1}{\sqrt{2}}\:
[\,V,\, V_{}^{\prime}]
\label{rg2}
\end{equation} 
where $H_{B}^{}$ and $H_{B_{}^{\prime}}^{}$ have the same DOS 
$\rho_{B}^{}$ and independent Poisson statistics, while 
$V$ and $V_{}^{\prime}$ both have the distribution defined in
 (\ref{xirand1}).
Clearly the number of energy levels  scales as 
$N \rightarrow 2 N$. 
We can choose for the  RG an iteration of the dimension 
doubling, hence a sequence
$\lambda_{k}^{}  = 2_{}^{- k/2} \lambda_{0}^{} \rightarrow 0$,
obtaining the  limit discussed in 
Section \ref{fluctuations}, when we  set $\varepsilon = 0$.

\section{More general models}
\label{modifiedmodel}

Here the solution of the model (\ref{twopart}) in Section  
\ref{simplemodel} is used to understand the properties of 
apparently more general models.
We will also be able to see the relation between the 
convolution property (\ref{cconv}) of the line shape function 
and the models defined here. 
Consider Hamiltonians of the form
 \begin{equation}
H_{1}^{}  = H_{0}^{} + V
\label{gen}
\end{equation}
where $H_{0}^{}$ is a diagonal matrix
with properties like those of $H_{B}^{}$ in (\ref{twopart}).
Again let $E_{k}^{}$ and  $\vert k \rangle$
be the eigenvalues and eigenvectors of $H_{B}^{}$, while the 
DOS is $\rho_{0}^{}$, and the spectral interval $\Delta$.
The  eigenvalues and eigenvectors of $H_{1}^{}$ are 
$\omega_{\nu}^{}$ and $\vert \omega_{\nu}^{}\rangle$. 
The matrix elements
\[ 
\langle j \vert V \vert k\rangle = 
\langle k  \vert  V \vert j \rangle_{}^{\ast}
= \xi_{jk}^{}
\]
are still assumed independent, identically distributed 
complex normal RVs
 \begin{equation}
\left \langle  \xi_{jk}^{} \right \rangle = 0, \qquad 
\left \langle  \xi_{j k}^{\ast}  \xi_{m n}^{}  \right \rangle 
= v_{}^{2} \delta_{j m}^{} \delta_{k n}^{} 
\label{vjk}
\end{equation} 
We note the known fact that the ensemble of such random matrices $V$ 
is invariant under all unitary transformations $U$, i.e. 
 $U_{}^{\dagger} V U  \sim V$ \cite{mehta}. 
Numerical simulations indicate that if 
$\Gamma := 2 \pi\rho_{0}^{} v_{}^{2}  \ll \Delta E $
and $\rho_{0}^{}\Gamma \gg 1$
then the DOS of $H_{1}^{}$ is also $\rho_{0}^{}$, 
except near the endpoints of $\Delta$.

Every eigenstate of $H_{0}^{}$ will decay as a result of the
interaction $V$.
Numerical evidence indicates that the decay rate is near $\Gamma$
when $\rho_{0}^{}\Gamma \gg 1$, and that the spectral densities 
are close to those found in Section  \ref{simplemodel}.
If this is true, then 
$\vert \langle k\vert \omega_{\nu}^{}\rangle \vert_{}^{2}$
inserted in (\ref{imr0}) gives a spectral density near the Lorentz form,
for every choice of $k$. (This fails for $E_{k}^{}$ near the end points
of $\Delta$, but we will argue as if  $\Delta = \R$.)
We now try to justify this picture, but without 
mathematical rigor.

Pick any eigenvalue of $H_{0}^{}$ that is not too close to the 
ends of $\Delta$. 
Call it $E_{0}^{}$  and  calculate how the corresponding eigenstate
$\vert 0 \rangle $ decays by  transforming this problem into that 
already solved in Section \ref{simplemodel}.
Make a decomposition   
\[
H_{1}^{} = H_{1A}^{} + H_{1B}^{} + V_{1}^{} + V_{1}^{\dagger}
\]
where there is only one non-zero element of 
$H_{1A}^{} = E_{0}^{} +\xi_{00}^{}$,   $V_{1}^{}$ is defined
by  
$(V_{1}^{})_{0k}^{} := \xi_{0k}^{} $ 
 and what is left  is $H_{1B}^{}$.
For each $\xi$  in the ensemble (\ref{vjk}) we can diagonalize 
$H_{1B}^{}$ by a unitary $U$ which leaves the  basis vector 
$\vert 0 \rangle$ invariant.
The transformed Hamiltonian is 
\[
U_{}^{\dagger} H_{1}^{} U = H_{1A}^{} 
+ U_{}^{\dagger} H_{1B}^{} U 
+ V_{1}^{} U + U_{}^{\dagger}V_{1}^{\dag}
\]
which is now of the  form  (\ref{twopart}),
where $H_{A}^{} \rightarrow H_{1A}^{}$ is 1-D, and 
$U_{}^{\dagger} H_{1B}^{} U$
is diagonal with eigenvalues 
randomly distributed, but with a level repulsion which makes 
the DOS nearly uniform and equal to $\rho_{0}^{}$ on the average.
Due to the unitary invariance of the ensemble (\ref{vjk}), 
it holds that $V_{1}^{} \sim V_{1}^{} U$.
Thus the solution in Section \ref{simplemodel}
will still hold here for the decay of the chosen state,
with $\rho_{B}^{} $ replaced by $\rho_{0}^{}$.
The fluctuation properties  will also
be the same.

We make the following note on the invariance properties of the 
ensemble  (\ref{vjk}).
The following phase transformation of the interaction part
\[
V_{jk}^{} \longrightarrow  
e_{}^{i(\theta_{j}^{} - \theta_{k}^{}  )}V_{jk}^{}
\]
leaves the ensemble invariant.
It is easy to check that the eigenvalues are invariant and 
that the eigenvectors are just multiplied by a phase
 \[ \langle k \vert \omega_{\nu}^{}\rangle 
\longrightarrow  e_{}^{ - i \theta_{k}^{}  }
\langle k \vert \omega_{\nu}^{}\rangle 
\]
Now let $\zeta_{k}^{} = \langle k \vert \zeta \rangle $ 
be the components of a vector. 
We can allow it to be a random vector, but assume it independent 
of the ensemble defined by (\ref{vjk}).
Then in 
\[
\langle \zeta \vert \omega_{\nu}^{}\rangle 
= \sum_{k}^{}\zeta_{k}^{}
\langle k \vert \omega_{\nu}^{}\rangle 
\] 
an average over
the ensemble  (\ref{vjk}), or just over all angles $\theta_{k}^{}$,
gives zero, while in the average of  the absolute square 
the cross terms vanish
\[
\left \langle \vert\langle \zeta \vert
 \omega_{\nu}^{}\rangle  \vert_{}^{2} \right \rangle 
= 
\sum_{k}^{} \vert \zeta_{k}^{}
\langle k \vert \omega_{\nu}^{}\rangle \vert_{}^{2}
\]
Using the spectral density form (\ref{imr0}) 
we find for $\rho_{0}^{} \Gamma  \gg 1$
\[
\left \langle \vert \langle \zeta \vert \omega_{\nu}^{} 
\rangle \vert_{}^{2} \right \rangle  
\approx \rho_{0}^{-1} 
\sum_{k}^{} \phi (\omega_{\nu}^{} - E_{k}^{}, \varepsilon)
 \vert  \zeta_{k}^{}  \vert_{}^{2}
\]
and replacing  $ \zeta_{k}^{} $
by a smooth function $\zeta( E_{k}^{})$ and approximating by
an integral
\[
\left \langle \vert \langle \zeta \vert \omega_{\nu}^{} 
\rangle \vert_{}^{2} \right \rangle  
\approx 
\int_{\R}^{} dx\, \phi  (\omega_{\nu}^{} -x, \varepsilon)
\,\vert \zeta( x)\vert_{}^{2}
\]
i.e. a convolution.

We now add another  interaction term with the
same statistics
\[
H_{2 }^{} = H_{1 }^{} + V_{}^{\prime}
=  H_{0}^{} + V + V_{}^{\prime}
\]
i.e. $ V \sim V_{}^{\prime}$, but we assume that $V$
and $V_{}^{\prime}$ are independent RVs.
By the rules of adding independent normal RVs with zero mean
$V + V_{}^{\prime} \sim \sqrt{2}\, V$.
The diagonalization of $H_{2 }^{}$ can be done in one 
step using the interaction $V + V_{}^{\prime}$, or in two,
first making $H_{1 }^{}$ diagonal, then $H_{2 }^{}$,   
while using the  unitary invariance of the ensembles.
Use   $\vert \Omega_{u}^{}\rangle $ for the eigenvectors 
of $H_{2 }^{}$. Then expand the scalar product
\[
\langle k \vert \Omega_{u}^{}\rangle 
= \sum_{\nu}^{}
\langle k \vert \omega_{\nu}^{}\rangle
\langle \omega_{\nu}^{}\vert \Omega_{u}^{}\rangle 
\]
take the absolute squared and a suitable ensemble average, as before 
\[
\left \langle \vert \langle k \vert \Omega_{u}^{}\rangle \vert_{}^{2}
\right \rangle
= \sum_{\nu}^{}
\vert\langle k \vert \omega_{\nu}^{}\rangle\vert_{}^{2}\:
\vert\langle \omega_{\nu}^{}\vert \Omega_{u}^{}\rangle \vert_{}^{2}
\]
If we average over the angles only we need no averages in the RHS.
Finally   with $\phi_{1}^{}$ the spectral density coming from 
diagonalizing $H_{1}^{}$ starting from $H_{0}^{}$, and  $\phi_{2}^{}$
from diagonalizing $H_{2}^{}$ starting from $H_{1}^{}$, 
we find the density for 
diagonalizing $H_{2}^{}$ starting from $H_{0}^{}$ as a convolution
\[
\phi(E_{k}^{} - \Omega_{u}^{}, 2 \varepsilon)
\approx \int dx \,
\phi_{1}^{}(E_{k}^{} - x,  \varepsilon) \,
\phi_{2}^{}(x - \Omega_{u}^{} ,  \varepsilon)
\]
With the explicit resonance form (\ref{imr2}), we find that
$\phi_{1}^{}$ and $\phi_{2}^{}$ have parameter $\Gamma$  
while that $\phi$ in the LHS has parameter
$2 \,\Gamma$, coming from $V + V_{}^{\prime} \sim \sqrt{2}\, V$.
Thus we  recover (\ref{cconv}) in the case
$\Gamma_{1}^{} =  \Gamma_{2}^{} = \Gamma$.

\section{Higher dimensions and Fano lineshapes}
\label{lineshape}

We return to the model (\ref{twopart}) and allow $H_{A}^{}$ to 
have a finite dimension $N_{A}^{}$. 
This kind of model can describe several states (interacting or
not), all decaying into quasicontinua. 
In such applications   we must make physical assumptions on the
matrix elements of $V$, and it is not always relevant 
to assume them all independent as in (\ref{vjk}).
It is easy to use two different quasicontinua with different
DOS, as in  (\ref{threepart}), perhaps coupled to two 
subset of levels in $H_{A}^{}$.

Pick one possible structure by choosing $\xi$ to be complex, 
normal RVs defined by 
\[
\langle \xi_{jr}^{} \rangle = 0,
\quad
\langle \xi_{jr}^{} \xi_{ks}^{\ast}  \rangle
= \gamma_{jk}^{}\: \delta_{rs}^{}
\]
where $\gamma$ is a positive semidefinite matrix of dimension
$N_{A}^{} \times N_{A}^{}$. 
The imaginary part of (\ref{htilde}) is then a positive 
semidefinite random matrix
 \begin{equation}
G_{jk}^{} := 
\left [\Im \tilde{H}_{A}^{}(x - i\varepsilon)\right]_{jk}^{} = 
\varepsilon \sum_{r}^{}  \frac{ \xi_{jr}^{} \xi_{kr}^{\ast}  }
{(x - E_{r}^{})_{}^{2} + \varepsilon_{}^{2}}
\end{equation} 
with ensemble average   
\[
\left \langle G_{jk}^{} \right\rangle 
= \gamma_{jk}^{} \: 
\varepsilon \,  \sum_{r}^{}
 \frac{1}{(x - E_{r}^{})_{}^{2} + \varepsilon_{}^{2}}
\approx \pi  \rho_{B}^{}\gamma_{jk}^{} 
\]
Comparing with the case $N_{A}^{} = 1$, it is natural to define
the golden rule matrix 
 \begin{equation}
\Gamma_{jk}^{}   := 2 \pi  \rho_{B}^{}\gamma_{jk}^{}  
\label{gr1}
\end{equation}
 The fluctuations around the average is estimated as in 
Section \ref{fluctuations}, and again found to  contain 
a factor $\omega_{B}^{} /\varepsilon$, just like 
(\ref{varh}).
Then (\ref{cscale})  scales  $\gamma$ as  
$\gamma \rightarrow \lambda_{}^{2}\gamma $,  
and we will find the convergence to the ensemble mean in the 
limit $\lambda \rightarrow 0$ just as for $N_{A}^{} = 1$.
The same holds for  $\Re \tilde{H}_{A}^{}$.
Consequently the exponential solution (\ref{ft}) 
(including $\varepsilon = 0$) is justified.

When $H_{A}^{}$ and $\Gamma$ commute and can be diagonalized 
simultaneously,  each eigenstate decays independently
and exponentially, each with a different Lorentz line shape 
and decay rate.
When they do not commute,  the present formalism 
includes models with  Fano line shapes \cite{fano61, durand01,dietz07}.
Assume that we can prepare the decaying system of $N_{A}^{}$
states in a definite quantum state $\vert \theta \rangle$.
The lineshape function which will govern the non-exponential
decay of this state is obtained from the 
spectral density  (\ref{imr1}) by projecting it on the 
corresponding 1-dimensional subspace
 (and setting $\varepsilon = 0$)
\[
f(\theta, x) :=\frac{1}{\pi}\, \langle \theta \vert  
\, \Im R_{A}^{}(x) \vert \theta \rangle 
\]
Clearly the factor $R_{A}^{}$ in  (\ref{imr1}) will 
in general have  $N_{A}^{}$ poles in the
upper half plane (and $R_{A}^{\dagger}$ the conjugate poles in the 
lower half plane); these poles will determine the spectral density, 
which is far from the simple Lorentz form in general.  
Depending on the vector $\vert \theta \rangle $  
there will be a weighted interference
of the contributions from the resonance poles. 
The standard Fano lineshapes can be reproduced for $N_{A}^{} = 2$ 
by suitable choices of the parameters. 
When $N_{A}^{} > 2$ the number of different possibilities
increase rapidly.

\section{Separation of time scales}
\label{timescales}

It is well known, for general models of decay, that if the energy
spectrum of the reservoir is bounded then the decay 
will be non-exponential for very short and for extremely long 
time scales; this subject was discussed in  e.g.\  \cite{fonda78}.
For the class of models considered here
the Lorentz line shape and exponential decay is an exact result in 
the limit (\ref{cscale}) if we also set the energy interval
$\Delta = \R$.
We will very briefly analyze  the deviations from this 
limit caused by a finite $\Delta$ or a finite $\rho_{B}^{}$.

First consider the case when $\Delta = \R$ but 
$\rho_{B}^{} < \infty$.
When $\{ E_{k}^{} = k \omega_{B}^{}; k \in \Z\}$, 
the ``decay'' is periodic with period 
$\tau_{B}^{} := 2 \pi \rho_{B}^{}$.
Clearly we need to have the decay time defined by $\Gamma$
to be much shorter than that. This is so if  
$\omega_{B}^{} \ll  \Gamma $, or, equivalently
$N_{\Gamma}^{} \gg 1$.
 However, in order to find the time we need to see the deviation
from an exponential decay to zero, another calculation is 
needed.
Recall that $\Phi_{A}^{} (x, 0)$ defined in (\ref{distr2})
is a step function reflecting the discrete spectrum, 
but it still has  a useful approximation 
(\ref{lim2}).
 The decay (characteristic) function  (\ref{char2})
will then be almost periodic. 
Consider the ergodic limit  
\[
\lim_{T \rightarrow \infty}^{} \frac{1}{T} 
\int_{0}^{T} dt\, \vert \chi_{A}^{}(t) \vert_{}^{2} 
\]
For exponentially decaying amplitudes this is zero,
but for  a discrete nondegenerate spectrum the limit picks 
out conjugate Fourier coefficients of 
$\chi_{A}^{}(t)$ and 
$\chi_{A}^{}(t)_{}^{\ast}$ 
reducing the double sum to a single (diagonal) sum which is 
constant in time 
 \[
\lim_{T \rightarrow \infty}^{} \frac{1}{T} 
\int_{0}^{T}dt\, \vert \chi_{A}^{}(t) \vert_{}^{2} 
\approx \frac{\omega_{B}^{2} \Gamma_{}^{2}}{(2\pi)_{}^{2}} 
\sum_{k}^{} \left [(E_{s}^{} - E_{k}^{})_{}^{2} 
+ \Gamma_{}^{2}/4 \right]_{}^{-2}
\]
When the sum is approximated by an integral we find that
 \[
\lim_{T \rightarrow \infty}^{} \frac{1}{T} 
\int_{0}^{T} dt\,\vert \chi_{A}^{}(t) \vert_{}^{2} 
\propto \frac{\omega_{B}^{}}{\Gamma} 
=  \frac{1}{N_{\Gamma}^{}}
\]
Thus, over   long times the function  
$\vert \chi_{A}^{}(t) \vert_{}^{2}$
is almost periodic, fluctuating around a non-zero mean value 
determined by $ N_{\Gamma}^{}$.
We can expect to see the exponential decay in this fluctuating 
background only for $t$ satisfying
\begin{equation}
 t \, \Gamma  < \ln N_{\Gamma}^{}
\label{long}
\end{equation}

Next we consider the effect of a finite $\Delta E$ while we 
can let  $\rho_{B}^{} = \infty$.
Then  $\vert \chi_{A}^{}(t) \vert_{}^{2}$ 
has a smooth quadratic maximum at $t = 0$.
Numerical simulations for large values of $\rho_{B}^{}$
(taking into account the level shift) 
show that this function is close to an exponential for 
\begin{equation}
t \, \Delta E  \gg 1
\label{short}
\end{equation} 
but they also indicate that there is a deviation from the exponential
\[
 \vert \chi(t) \vert_{}^{2} - \exp( - \Gamma t)
  = O( \Gamma/ \Delta E)
\]which remains significant over a time interval of the 
order of $1/\Gamma$.
When $\Gamma \ll \Delta E$ this deviation is very small, and the time 
$1/\Delta E$ is very short compared to the relaxation time scale.
Then the effect of a finite
$\Delta E $ on the relaxation will be insignificant, and the 
exponential is a good approximation.

\section{Conclusions}
\label{discussion}

In this paper we argue that the exponential relaxation 
and golden rule for model Hamiltonians of the type 
(\ref{twopart})  can be understood as fixed point properties 
of this class under a simple RG of transformations, 
e.g. in the form (\ref{rg1}). 

The mathematical basis for the results is the simple 
non-perturbative form (\ref{imr1}) for the spectral density 
and the assumed uniformity of the spectrum of $H_{B}^{}$
and the matrix elements of $V$, for instance in terms of 
the relations (\ref{uni1}) and (\ref{uni2}).
In the random matrix approach the results hold with probability
one for each element in the ensemble.
The method works for  a single  decaying state, with a simple 
Lorentzian lineshape, as well  as   more complex cases, including 
Fano lineshapes,  and it does not involve an expansion in the coupling
strength.  
Instead we estimated the deviations from the  scaling limit  when
the scaling parameter $\lambda$ is small but  nonzero, 
and found that the deviation is small
if the number of states under the resonance is large: 
$ N_{\Gamma}^{} = \rho_{B}^{} \Gamma\gg 1  $.
We could also handle the case of a   finite energy range $\Delta E $
for $H_{B}^{}$. Using the dimensionless quantities $N_{\Gamma}^{}$
and $N =  \rho_{B}^{}\Delta E $ (the total number of states) 
the condition for small deviations 
(\ref{ineq}) reads
\[
1 \ll N_{\Gamma}^{} \ll N 
\]
In Section \ref{timescales} we concluded that the exponential decay 
is a very good  approximation in a time interval restricted by 
 (\ref{long}) and (\ref{short}). 
In terms of $N$ and $N_{\Gamma}^{}$ we found that 
\[
N_{\Gamma}^{}/ N \ll t \, \Gamma  < \ln N_{\Gamma}^{}
\]

On the other hand, we note that these models cannot be used to derive 
the most general irreversible evolution for open quantum systems.
The CF  (\ref{ft}), acting on any pure initial state in the subspace $A$, 
defines a pure state, with decreasing norm,  for all $t > 0$. 
This is clearly a special case, but still useful in many applications.

In view of the frequent use of the van Hove limit (\ref{wscale})
it is important to say that 
(\ref{cscale}) is  not a weak coupling limit;
instead it must be interpreted as a limit where the strength
of the coupling is held constant while the reservoir is enlarged.
In order to see this we first have to find a  measure
of the strength of the 
interaction term in the model of Section \ref{simplemodel}.
When  $N_{B}^{} < \infty$ then we can choose the square root 
of the positive scalar (expectation 
for a random $V$) 
\[
\left \langle V V_{}^{\dag} \right \rangle 
 = N_{B}^{} v_{}^{2} \approx \rho_{B}^{}  v_{}^{2} \Delta E
= \frac{1}{2 \pi}\:\Gamma \Delta E
\]  
to get a measure of dimension energy.
In the scaling (\ref{cscale}) the parameters $\Gamma$ 
and $\Delta E $ are invariant, so is the strength of the 
interaction. 
When $\Delta E \rightarrow \infty$ this value
is inevitably misleading. 
The matrix elements connecting the decaying level to 
off-resonant final states, i.e. with 
$ \vert E_{s}^{} - E_{k}^{} \vert \gg \Gamma $, 
will have little influence on the decay except for very short 
or very long times. 
Provided $\Gamma \ll \Delta E$ 
we can then replace $\Delta E$ by a multiple of $\Gamma$ 
and use the scaling invariant $\Gamma$ itself as a useful measure 
of the strength of the interaction.

\subsection*{Acknowledgments}
This work was supported by the Swedish Science Research Council (VR).


\end{document}